\documentclass[particles,article,accept,pdftex,moreauthors]{Definitions/mdpi}
\firstpage{1}
\pubvolume{1}
\issuenum{1}
\articlenumber{0}
\pubyear{2026}
\copyrightyear{2026}
\datereceived{ }
\daterevised{ } % Comment out if no revised date
\dateaccepted{ }
\datepublished{ }
\usepackage{bm}
\newif\ifshowrevisions
\showrevisionstrue
\ifshowrevisions
  \usepackage[normalem]{ulem}
\Title{Effects of Isovector Spin--Orbit Interaction on the Charge-Weak Form Factor Difference in $^{48}$Ca, $^{208}$Pb, $^{90}$Zr and $^{62}$Ni}

\Author{Tong-Gang Yue %MDPI: Please carefully check the accuracy of names and affiliations.
 $^{1,2}$, Zhen Zhang $^{3,}$*\orcidB{} and Lie-Wen Chen $^{1,}$*\orcidC{}}

\AuthorNames{Tong-Gang Yue, Zhen Zhang and Lie-Wen Chen}

\address{%
$^{1}$ \quad State Key Laboratory %MDPI: There are three Laboratory in this affiliation. If the information provided contains more than one address, please separate the addresses into different affiliations.
 of Dark Matter Physics, Key Laboratory for Particle Astrophysics and Cosmology (MOE), and Shanghai Key Laboratory for Particle Physics and Cosmology, School of Physics and Astronomy, Shanghai Jiao Tong University, Shanghai 200240, China; 1076788852@sjtu.edu.cn %MDPI: We added the email address here according to that submitted online at susy.mdpi.com. Please confirm.
\\
$^{2}$ \quad Tsung-Dao Lee Institute, Shanghai Jiao Tong University, Shanghai 201210, China\\
$^{3}$ \quad Sino-French Institute of Nuclear Engineering and Technology, Sun Yat-sen University, Zhuhai 519082, China}

\corres{Correspondence: zhangzh275@mail.sysu.edu.cn %MDPI: The email address is different from the one submitted online at susy.mdpi.com. Please confirm which is correct.
 (Z.Z.); lwchen@sjtu.edu.cn (L.-W.C.)}

\abstract{The nucleon spin–orbit interaction is a cornerstone of nuclear structure theory, yet its isospin dependence remains {insufficiently constrained within modern nuclear energy density functional (EDF) theory. It was recently shown that, within the framework of extended Skyrme EDFs, the charge-weak form factor difference $\Delta F_{\rm CW}$ in $^{48}$Ca exhibits remarkable sensitivity to the effective isovector spin–orbit (IVSO) interaction, whereas $\Delta F_{\rm CW}$ in $^{208}$Pb is much less sensitive to this channel.} Extending this analysis to other nuclei, we find that $^{90}$Zr, with its ten spin–orbit unpaired $1\mathrm{g}_{9/2}$ neutrons, displays a $\Delta F_{\rm CW}$ sensitivity to the IVSO strength similar to that of $^{48}$Ca, arising from modifications to the central mean-field potential rather than the one-body spin–orbit potential. In contrast, $^{62}$Ni, like $^{208}$Pb, remains largely insensitive to the IVSO interaction. This structure-driven distinction suggests an experimental strategy: future parity-violating electron scattering measurements, e.g., the MREX experiment at the MESA facility, on $^{48}$Ca and $^{90}$Zr would help constrain the effective IVSO strength, while measurements on $^{208}$Pb and $^{62}$Ni can provide a cleaner probe of the density dependence
of the symmetry energy with reduced IVSO sensitivity.}


\begin{document}

%%%%%%%%%%%%%%%%%%%%%%%%%%%%%%%%%%%%%%%%%%
\section{Introduction}

The spin–orbit (SO) interaction plays a crucial role across many branches of physics---from elementary quark--gluon dynamics~\cite{Liang:2004ph,STAR:2017ckg} to macroscopic phenomena in condensed matter and optics~\cite{Bihlmayer2022,Bliokh:2015yhi}---and has long been recognized as a fundamental ingredient of nuclear structure~\cite{Thomas:1926dy,Dirac:1928hu,Kaiser:1997mw,Kaiser:2003ff}. Historically, it was first introduced in atomic physics in the 1920s to explain the fine structure of atomic spectra~\cite{Goudschmidt:1926ea,Thomas:1926dy}.
In 1949, Mayer~\cite{Mayer:1949pd} and Jensen \cite{Haxel:1949fjd} incorporated a strong nucleon SO interaction into the nuclear shell model, successfully explaining the emergence of magic numbers in nuclei. This achievement marked a milestone in nuclear theory, establishing SO interaction as a key mechanism for understanding nuclear structure, stability, and shell evolution.

Since nuclei are composed of protons and neutrons, a natural question then arises: is the SO interaction identical for protons and neutrons in the nuclei? One of its most direct manifestations is the spin–orbit splitting, namely, the splitting of a single-particle level with a given orbital angular momentum $l$ ($l\neq 0$) into two sublevels with different total angular momenta $j=l\pm 1/2$. For nuclei close to the $\beta$-stability line, the isospin asymmetry (measuring difference between neutron and proton numbers) is small, leading to similar magnitudes of spin–orbit splitting for protons and neutrons. In contrast, the nuclei far from stability, particularly those with a large neutron excess, exhibit marked differences between proton and neutron spin–orbit splittings~\cite{Isakov:2002jv}. This observation suggests the isospin-dependent nature of the SO interaction, i.e., an isovector spin–orbit~(IVSO) interaction. However, since the single-nucleon energy levels cannot be directly measured and the impact of SO interaction has a complex interplay with, e.g., tensor force~\cite{Lesinski:2007ys} and nucleon effective masses~\cite{Klupfel:2008af}, the quantitative strength of the IVSO channel remains model dependent and insufficiently constrained at the energy-density-functional~(EDF) level~\cite{Bender:2003jk}.

{In modern nuclear EDF theory, the origin and predictive content of spin–orbit couplings depend on how the functional is constructed. In covariant mean-field models, a strong SO potential emerges naturally from the Dirac structure of the scalar and vector mean fields, while the isospin dependence is weak in conventional $\sigma$-$\omega$-$\rho$ parametrizations~\cite{Sharma:1994mim,Reinhard:1995}. Additional isovector degrees of freedom, such as the isovector-scalar $\delta$ meson and exchange terms in covariant EDFs, can modify the effective isospin dependence of the SO channel~\cite{Ebran:2016rjd}. In nonrelativistic Skyrme EDFs, the SO term is usually introduced phenomenologically~\cite{Bender:2003jk}, and the conventional zero-range two-body form gives an IVSO strength only one third of the isoscalar one~\cite{Vautherin:1971aw}. A different baseline is provided by EDFs constrained by pion-nucleon dynamics, where the analytical form and density-dependent couplings are derived from the many-body perturbation theory and few-nucleon interactions~\cite{Kaiser:2010pp,Holt:2011nj}. These calculations generate nontrivial density-dependent SO couplings and predict an IVSO strength about twice as large as that in conventional Skyrme EDFs~\cite{Kaiser:2010pp}. Subsequent studies of chiral forces found a sizable enhancement of the isoscalar SO interaction~\cite{Holt:2011nj} due to the three-body contributions but did not provide a corresponding analysis of the IVSO channel. This model dependence highlights the need for effective experimental probes of the IVSO~channel.}

Recent studies have identified the charge-weak form factor difference $\Delta F_{\rm CW}$ in ${}^{48}$Ca as a sensitive probe of the IVSO interaction~\cite{Yue:2024srj,Zhao:2024gjz,Kunjipurayil:2025xss}. In particular, the model-independent determination of $\Delta F_{\rm CW}$ in ${}^{48}$Ca by the Calcium Radius Experiment (CREX)~\cite{CREX:2022kgg} at Jefferson Lab (JLab), together with the PREX-II result in ${}^{208}$Pb~\cite{PREX:2021umo}, has highlighted the difficulty of accommodating both measurements simultaneously within conventional EDFs~\cite{Reinhard:2022inh,Yuksel:2022umn,Zhang:2022bni,Mondal:2022cva,Tagami:2022spb,Miyatsu:2023lki,Sammarruca:2023mxp,Papakonstantinou:2022gkt,Kumar:2023bmb,Lattimer:2023rpe,Liliani:2023cfv} as well as in ab initio calculations~\cite{CREX:2022kgg,Hu:2021trw}. Within the extended Skyrme EDF framework, a stronger IVSO channel has been found to improve the simultaneous accommodation of the CREX and PREX-II $\Delta F_{\rm CW}$ data, while also highlighting the distinct sensitivities of different nuclei to the IVSO interaction~\cite{Yue:2024srj}. Very recently, a study based on covariant density-dependent point-coupling EDFs with an isovector tensor coupling reported a similar phenomenological trend~\cite{Qiu:2025qbv}.

In this work, through an analysis of the single-particle levels in $^{48}\mathrm{Ca}$ and $^{208}\mathrm{Pb}$, we demonstrate that the conventional shell closures, including the magic numbers, are well reproduced and retain their robustness even in the presence of a strong IVSO interaction. Extending this analysis, we identify a structural correspondence between $^{90}\mathrm{Zr}$ and $^{48}\mathrm{Ca}$, as well as between $^{62}\mathrm{Ni}$ and $^{208}\mathrm{Pb}$, based on their analogous spin–orbit saturation properties. This analogy implies that the charge-weak form factor difference $\Delta F_{\mathrm{CW}} = F_C - F_W$, and thereby the neutron skin, exhibits significant sensitivity to the IVSO strength in $^{90}\mathrm{Zr}$ while remaining largely insensitive in $^{62}\mathrm{Ni}$. The broader implications of this nucleus-dependent sensitivity are discussed in what follows.

%%%%%%%%%%%%%%%%%%%%%%%%%%%%%%%%%%%%%%%%%%
\section{Methods}
\subsection{Extended Skyrme EDF Within Hartree-Fock-Bogoliubov (HFB) Framework}

In the present work, we use a nonrelativistic nuclear EDF based on an extended Skyrme interaction~\cite{Chamel:2009yx} supplemented by a zero-range tensor force~\cite{Stancu:1977va}. The extended Skyrme interaction $v_{eSky}$ differs from the conventional Skyrme interaction $v_{Sky}$ by including momentum-dependent many-body terms $v^{\prime}$, namely,\vspace{-3pt}
\begin{eqnarray}
v_{eSky}=v_{Sky}+v^{\prime}.
\end{eqnarray}
The %MDPI: Please confirm whether the paragraph below the formula needs to be indented. The same to following highlight without comment.
 full interaction used in this work consists of the extended Skyrme part $v_{eSky}$ together with the zero-range tensor force $V_T$ introduced below.

The conventional Skyrme interaction $v_{Sky}$ reads (see, e.g., Ref.~\cite{Chabanat:1997qh}): %MDPI: Please ensure all variables/values in the equation appear in the same format in the text (normal/italic/bold). Please check and revise the entire text.
\vspace{-3pt}
\begin{eqnarray}
\begin{aligned}
& v_{Sky}=t_0\left(1+x_0 P_\sigma\right) \delta(\bm{r}) +\frac{1}{2} t_1\left(1+x_1 P_\sigma\right)\left[\bm{k}^{\prime 2} \delta(\bm{r})+\delta(\bm{r}) \bm{k}^2\right] \\
& +t_2\left(1+x_2 P_\sigma\right) \bm{k}^{\prime} \cdot \delta(\bm{r}) \bm{k} +\frac{1}{6} t_3\left(1+x_3 P_\sigma\right)[\rho(\bm{R})]^\alpha \delta(\bm{r}) \\
&+ \mathrm{i} W_0 \left(\hat{\bm{\sigma}}_1+\hat{\bm{\sigma}}_2\right)\cdot\left[\bm{k}^{\prime} \times \delta(\bm{r})\,\bm{k}\right],
%ZZ: We have highlighted the revised notation to keep vector and imaginary-unit formatting consistent.
\end{aligned}
\end{eqnarray}
and\vspace{-3pt}
\begin{eqnarray}
\begin{aligned}
v^{\prime} &= \frac{1}{2} t_4\left(1+x_4 P_\sigma\right)\left[\bm{k}^{\prime 2} \rho(\bm{R})^\beta \delta(\bm{r})+\delta(\bm{r}) \rho(\bm{R})^\beta \bm{k}^2\right] \\
&+ t_5\left(1+x_5 P_\sigma\right) \bm{k}^{\prime} \cdot \rho(\bm{R})^\gamma \delta(\bm{r}) \bm{k}.
\end{aligned}
\end{eqnarray}
Here, $\bm{r}=\bm{r}_1-\bm{r}_2$, $\bm{R}=\frac{1}{2}\left(\bm{r}_1+\bm{r}_2\right)$, $\bm{k}=\frac{1}{2\mathrm{i}}\left(\bm{\nabla}_1-\bm{\nabla}_2\right)$ is the relative momentum, with $\bm{k}^{\prime}$ being its conjugate acting on the left; the $\hat{\bm{\sigma}}$ are Pauli spin operators, $P_{\sigma}=\frac{1}{2}\left(1+\hat{\bm{\sigma}}_1 \cdot \hat{\bm{\sigma}}_2\right)$ is the spin exchange operator,
and $\rho(\bm{R})=\rho_n(\bm{R})+\rho_p(\bm{R})$ is the total local density, with $\rho_n$ and $\rho_p$ being neutron and proton densities, respectively. $t_0\sim t_5$, $x_0\sim x_5$, $\alpha$, $\beta$ and $\gamma$ are adjustable ``Skyrme parameters'', and $W_0$ is the SO interaction parameter. For simplicity, both $\beta$ and $\gamma$ are set to unity as in Ref.~\cite{Zhang:2015vaa}%Please check intended meaning has been retained.
.

For the tensor force, we use the following zero-range form~\cite{Stancu:1977va}\vspace{-3pt}
\begin{equation}
\begin{aligned}
V_T &= \frac{1}{2} T \biggl\{
        \left[
            \left(\hat{\bm{\sigma}}_1 \cdot \bm{k}^{\prime}\right)
            \left(\hat{\bm{\sigma}}_2 \cdot \bm{k}^{\prime}\right)
            - \frac{1}{3} \bm{k}^{\prime 2}
            \left(\hat{\bm{\sigma}}_1 \cdot \hat{\bm{\sigma}}_2\right)
        \right] \delta\left(\bm{r}\right) \\
    &\quad + \delta\left(\bm{r}\right)
        \left[
            \left(\hat{\bm{\sigma}}_1 \cdot \bm{k}\right)
            \left(\hat{\bm{\sigma}}_2 \cdot \bm{k}\right)
            - \frac{1}{3} \bm{k}^2
            \left(\hat{\bm{\sigma}}_1 \cdot \hat{\bm{\sigma}}_2\right)
        \right]
    \biggr\} \\
    &+ U \biggl\{
        \left(\hat{\bm{\sigma}}_1 \cdot \bm{k}^{\prime}\right)
        \delta\left(\bm{r}\right)
        \left(\hat{\bm{\sigma}}_2 \cdot \bm{k}\right) - \frac{1}{3}
        \left(\hat{\bm{\sigma}}_1 \cdot \hat{\bm{\sigma}}_2\right)
        \left[
            \bm{k}^{\prime} \cdot \delta\left(\bm{r}\right) \bm{k}
        \right]
    \biggr\},
\end{aligned}\vspace{-6pt}
\end{equation}
where $T$ and $U$ quantify the magnitude of the tensor force in states of even and odd relative motion, respectively. In practical calculations, the parameters $\alpha_T$ and $\beta_T$ are employed in place of $U$ and $T$, where $\alpha_T$ and $\beta_T$ are expressed as:
\begin{equation}
\alpha_T = \frac{5}{12} U, \quad \beta_T = \frac{5}{24} (T + U).
\end{equation}
The explicit zero-range tensor force contributes to the EDF through the $\alpha_T$- and $\beta_T$-dependent $\bm{J}^2$ and $\tilde{\bm{J}}^2$ terms in Equation~(\ref{Eq:Sky-HF}).

In addition, for calculations of open-shell nuclei, we take into account the standard mixed volume-surface density-dependent contact pairing force of the form~\cite{Dobaczewski:2002mt}
\begin{equation}
V_p(\bm{r},\bm{r}^{\prime}) =  V_0 \left[ 1 - \frac{\rho(\bm{r})}{2\rho_0} \right] \delta(\bm{r}-\bm{r}^{\prime}),
\end{equation}
where $V_0$ is the pairing strength and $\rho_0 = 0.16~\rm fm^{-3}$.

Based on the above extended Skyrme interaction, tensor force and pairing force, we derive the EDF within the HFB framework. Under the assumption of time-reversal invariance, the energy of a nucleus can be written
as the integral of a purely local EDF $\mathcal{E}$. For details on the HFB calculation based on Skyrme EDF, see e.g., Ref.~\cite{Stoitsov:2004pe}. In the following, we focus on the potential-energy density $\mathcal{E}_{eSky}$ associated with the extended Skyrme interaction and tensor force, which can be expressed in terms of
local density $\rho_t$, kinetic-energy density $\tau_t$, and vector spin-current density $\bm{J}_t$ of neutrons ($t=n$) and protons ($t=p$) defined by\vspace{-4pt}
\begin{eqnarray}
        \rho_{t}(\bm{r}) & =&\sum_{i} v_i^2\left|\varphi_i(\bm{r})\right|^2, \\
        \tau_{t}(\bm{r}) & =&\sum_i v_i^2\left|\bm{\nabla} \varphi_i(\bm{r})\right|^2, \\
        \bm{J}_{t}(\bm{r}) & =&-\mathrm{i} \sum_i v_i^2 \varphi_i^{+}(\bm{r}) \bm{\nabla} \times \hat{\bm{\sigma}} \varphi_i(\bm{r}).
\end{eqnarray}
Here, $\varphi_i$ is the wave function of canonical state $i$, $v_i$ is the corresponding occupation probability, and the sum runs over all states for the same species $t$. By further defining the isoscalar densities
$$ \rho =\rho_n+\rho_p,~~\tau = \tau_n+\tau_p, ~~\bm{J}= \bm{J}_n+\bm{J}_p,$$
and isovector densities
$$ \tilde{\rho} =\rho_n-\rho_p,~~\tilde{\tau} = \tau_n-\tau_p, ~~\tilde{\bm{J}}= \bm{J}_n-\bm{J}_p,$$
the $\mathcal{E}_{eSky}$ can be written as\vspace{-3pt}
\begin{eqnarray} \label{Eq:Sky-HF}
    \mathcal{E}_{eSky} &=&
\frac{B_0+B_3 \rho^\alpha}{2} \rho^2-\frac{B_0^{\prime}+B_3^{\prime} \rho^\alpha}{2} \tilde{\rho}^2+(B_1+B_4\rho^{\beta}+B_5\rho^{\gamma})\rho \tau \notag\\
&&-(B_1^{\prime}+B_4^{\prime}\rho^{\beta}+B_5^{\prime}\rho^{\gamma}) \tilde{\rho} \tilde{\tau}
+\frac{2B_2+(2\beta+3)B_4\rho^{\beta}-B_5\rho^{\gamma}}{4} (\bm{\nabla} \rho)^2 \notag \\
&&-\frac{2B_2^{\prime}+3B_4^{\prime}\rho^{\beta}-B_5^{\prime}\rho^{\gamma}}{4}(\bm{\nabla}{\tilde{\rho}})^2
-\frac{\beta B_4^{\prime}}{2}\rho^{\beta-1}\tilde{\rho}\bm{\nabla} \rho \cdot \bm{\nabla} \tilde{\rho}
 \notag\\
&&+\frac{C_1+C_2\rho^{\beta}+C_3\rho^{\gamma}}{2} \bm{J}^2 + \frac{C_1^{\prime}+C_2^{\prime} \rho^{\beta} + C_3^{\prime} \rho^{\gamma} }{2} \tilde{\bm{J}}^2  \notag \\
        &&+\frac{b_{\rm{IS}}}{2} \bm{\nabla} \rho \cdot \bm{J} + \frac{b_{\rm{IV}}}{2} \bm{\nabla} \tilde{\rho} \cdot \tilde{\bm{J}}+\frac{\alpha_T + \beta_T}{4} \bm{J}^2 + \frac{\alpha_T - \beta_T}{4} \tilde{\bm{J}}^2 .
\end{eqnarray}

The coefficients $B_i$, $B_i^{\prime}$, $C_i$ and $C_i^{\prime}$ are uniquely related to the Skyrme parameters $t_i$ and $x_i$ by
{\small
\begin{align}
\label{Eq:BC}
B_0 &= \frac{3}{4}t_0, &
B_0^{\prime} &= \frac{1}{2}t_0 \left(\frac{1}{2} + x_0\right), \notag \\
B_1 &= \frac{3}{16} t_1 + \frac{5}{16} t_2 + \frac{1}{4} t_2 x_2, &
B_1^{\prime} &= \frac{1}{8}\left[t_1\left(\frac{1}{2}+x_1\right)-t_2\left(\frac{1}{2}+x_2\right)\right], \notag \\
B_2 &= \frac{9}{32} t_1 - \frac{5}{32} t_2 - \frac{1}{8} t_2 x_2, &
B_2^{\prime} &= \frac{1}{16}\left[3 t_1\left(\frac{1}{2}+x_1\right)+t_2\left(\frac{1}{2}+x_2\right)\right], \notag \\
B_3 &= \frac{1}{8} t_3, &
B_3^{\prime} &= \frac{1}{12} t_3 \left(\frac{1}{2} + x_3\right), \notag \\
B_4 &= \frac{3}{16} t_4, &
B_4^{\prime} &= \frac{1}{8}t_4\left(\frac{1}{2}+x_4\right), \notag \\
B_5 &= \frac{5}{16} t_5 + \frac{1}{4} t_5 x_5, &
B_5^{\prime} &= -\frac{1}{8}t_5\left(\frac{1}{2}+x_5\right), \notag \\
C_1 &= \eta_{\mathrm{tls}} \frac{1}{8}\left[t_1\left(\frac{1}{2}-x_1\right)-t_2\left(\frac{1}{2}+x_2\right)\right], &
C_1^{\prime} &= \eta_{\mathrm{tls}} \frac{1}{16}(t_1 - t_2), \notag \\
C_2 &= \eta_{\mathrm{tls}} \frac{1}{8}t_4\left(\frac{1}{2}-x_4\right), &
C_2^{\prime} &= \eta_{\mathrm{tls}} \frac{1}{16} t_4, \notag \\
C_3 &= -\eta_{\mathrm{tls}} \frac{1}{8}t_5\left(\frac{1}{2}+x_5\right), &
C_3^{\prime} &= -\eta_{\mathrm{tls}} \frac{1}{16} t_5.
\end{align}}The $b_{\rm IS}$ and $b_{\rm IV}$ characterize the isoscalar and isovector SO strengths, respectively. For the standard zero-range two-body SO term in the Skyrme interaction, one has\linebreak   $b_{\rm IS}=3b_{\rm IV}=3W_0/2$, where $W_0$ is the Skyrme SO parameter.  In the present work, however, following Ref.~\cite{Reinhard:1995}, $b_{\rm IS}$ and $b_{\rm IV}$ are treated as independent phenomenological EDF-level couplings, rather than as parameters of a strict short-range two-body SO interaction, in order to explore the sensitivity of finite-nucleus observables to the IVSO channel within the adopted EDF framework.

Following Ref.~\cite{Yue:2024srj}, we employ the eS53 and eS250 Skyrme EDFs constructed therein. Among them, eS250 is a fully optimized strong-IVSO functional with $(b_{\rm IS},b_{\rm IV})\approx(160,250)$ MeV$\cdot$fm$^5$. As shown in Ref.~\cite{Yue:2024srj}, eS250 {gives $\Delta F_{\rm CW}$ predictions for $^{48}$Ca and $^{208}$Pb that are consistent with the CREX and PREX-II data within their respective $1\sigma$ uncertainties}. The functional eS53, with $(b_{\rm IS},b_{\rm IV})\approx(160,53)$ MeV$\cdot$fm$^5$ corresponding to the conventional relation $b_{\rm IV}=b_{\rm IS}/3$, is obtained from eS250 by modifying the IVSO strength while keeping the other EDF parameters unchanged. For reference, we also discuss the reference functional eS500$_{\rm T}$ from Ref.~\cite{Yue:2024srj}, which is likewise fully optimized and adopts a very strong IVSO coupling with $(b_{\rm IS},b_{\rm IV})\approx(134,500)$ MeV$\cdot$fm$^5$.

In Equation~(\ref{Eq:Sky-HF}), the coefficients $C_i$ and $C_i^{\prime}$ collect the $\bm{J}^2$ and $\tilde{\bm{J}}^2$ contributions associated with the momentum-dependent central part of the extended Skyrme interaction, whereas the zero-range tensor force contributes separately through the $\alpha_T$- and $\beta_T$-dependent terms.
In the EDF implementation, the parameter $\eta_{\rm tls}$ is used as a bookkeeping switch to control whether the central contribution to the $\bm{J}^2$ sector is retained~\cite{Klupfel:2008af}. For the EDFs employed in this work, $\eta_{\rm tls}=0$ for both eS53 and eS250, whereas $\eta_{\rm tls}=1$ for eS500$_{\rm T}$.
%ZZ: We have highlighted the revised spin-current notation for consistency.

Based on the above EDF, nuclear ground-state properties are calculated by solving the HFB equations. The single nucleon Hamiltonian entering into the HFB equation can be expressed as\vspace{-4pt}
\begin{eqnarray}\label{Eq:hq}
\hat{h}_{t} =-\bm{\nabla} \cdot \frac{\hbar^2}{2m_{t}^*}{\bm{\nabla}} +U_{t}+\mathrm{i}\bm{W}_{t}\cdot(\hat{\bm{\sigma}}\times \bm{\nabla}),~ t=n,~p,
\end{eqnarray}
where the single-nucleon fields can be derived from the energy density $\mathcal{E}$ by\vspace{-4pt}
\begin{equation}
\begin{aligned}
\frac{\hbar^2}{2 m_{t}^*} & =\frac{\partial \mathcal{E}}{\partial \tau_{t}}, ~~U_{t}=\frac{\partial \mathcal{E}}{\partial \rho_{t}}-\bm{\nabla} \cdot \frac{\partial \mathcal{E}}{\partial\left[\bm{\nabla} \rho_{t}\right]},~~
        \bm{W}_{t} & =\frac{\partial \mathcal{E}}{\partial \bm{J}_{t}} .
\end{aligned}
\end{equation}
From Equation~(\ref{Eq:Sky-HF}), one can obtain the nucleon effective mass $m_t^*$, i.e.,\vspace{-4pt}
\begin{eqnarray}
\frac{\hbar^2}{2m_{t}^*} &=& \frac{\hbar^2}{2m} +(B_1+B_4\rho^{\beta}+B_5\rho^{\gamma})\rho -\eta_t(B_1^{\prime}+B_4^{\prime}\rho^{\beta}+B_5^{\prime}\rho^{\gamma}) \tilde{\rho},
\end{eqnarray}
the single nucleon potential (without Coulomb potential)
\begin{adjustwidth}{-\extralength}{0cm}
\vspace{-13pt}
\begin{eqnarray}
U_{t} &=& B_0\rho -\eta_t B_0^{\prime}\tilde{\rho}+B_1\tau -\eta_t B_1^{\prime}\tilde{\tau}+\frac{\alpha+2}{2}B_3 \rho^{\alpha+1} -\frac{B_3^{\prime}}{2}(\alpha \tilde{\rho} +2\eta_t \rho)\rho^{\alpha-1}\tilde{\rho} \notag \\
&&+(\beta+1)B_4\rho^{\beta}\tau+(\gamma+1)B_5\rho^{\gamma}\tau-(B_4^{\prime}\beta \rho^{\beta}+B_5^{\prime}\gamma \rho^{\gamma})\tilde{\rho}\tilde{\tau}-\eta_t(B_1^{\prime}+B_4^{\prime}\rho^{\beta}+B_5^{\prime}\rho^{\gamma})\tilde{\tau} \notag \\
 &&-\frac{\beta(2\beta+3)B_4 \rho^{\beta-1}-B_5\gamma \rho^{\gamma-1}}{4}(\bm{\nabla} \rho)^2 -\frac{2B_2+(2\beta+3)B_4\rho^{\beta} -B_5\rho^{\gamma}}{2}\bm{\nabla}^2 \rho \notag \\
 && -\frac{3\beta B_4^{\prime}\rho^{\beta-1}-\gamma B_5^{\prime}\rho^{\gamma-1}}{4} \bm{\nabla} \tilde{\rho} \cdot(2\eta_t\bm{\nabla}\rho +\bm{\nabla} \tilde{\rho})-\frac{2B_2^{\prime}+3B_4^{\prime}\rho^{\beta}-B_5^{\prime}\rho^{\gamma}}{2}\eta_t\bm{\nabla}^2 \tilde{\rho}\notag \notag \\
 &&+\frac{\beta B_4^{\prime}}{2}\rho^{\beta-1}(\bm{\nabla} \tilde{\rho})^2 + \frac{\beta B_4^{\prime}}{2}\rho^{\beta-1}\tilde{\rho}\bm{\nabla}^2 \tilde{\rho}+ \frac{\beta(\beta-1)B_4^{\prime}}{2}\rho^{\beta-2}\tilde{\rho}(\bm{\nabla}\rho)^2\eta_t\notag \\
 &&+\frac{\beta B_4^{\prime}}{2}\rho^{\beta-1}\tilde{\rho}(\bm{\nabla}^2\rho)\eta_t+\frac{\beta C_2\rho^{\beta-1}+\gamma C_3\rho^{\gamma-1}}{2} \bm{J}^2+\frac{\beta C_2^{\prime}\rho^{\beta-1}+\gamma C_3^{\prime}\rho^{\gamma-1}}{2} \tilde{\bm{J}}^2\notag\\
&&-\frac{b_{\rm{IS}}}{2}\bm{\nabla} \cdot \bm{J}-\eta_t\frac{b_{\rm{IV}}}{2}\bm{\nabla} \cdot \tilde{\bm{J}},
%& &+\frac{\beta^2 B_4^{\prime}}{2}\eta_t(\nabla \rho)^2 +\frac{\beta B_4^{\prime}}{2}\rho^{\beta} \eta_t\nabla^2 \rho \\
\label{Eq:CPot}
\end{eqnarray}
\end{adjustwidth}
and the SO potential
\begin{eqnarray}
\bm{W}_{t}&=& \frac{b_{\rm IS}}{2}\bm{\nabla}
\rho+\eta_t \frac{b_{\rm IV}}{2}\bm{\nabla}(\rho_n-\rho_p) + \frac{\alpha_{J}+\beta_{J}}{2}\bm{J} + \eta_t \frac{\alpha_{J}-\beta_{J}}{2}(\bm{J}_n-\bm{J}_p),
%ZZ: We have highlighted the revised roman subscripts for consistency with the surrounding text.
\label{Eq:SOPot}
\end{eqnarray}
where $\eta_t = 1$ and $-1$ for $t= n$ and $p$, respectively. The parameters  $\alpha_{J} $ and $\beta_{J}$ in the SO potentials are defined by
\begin{eqnarray}
\alpha_{J} = \alpha_{C}+\alpha_{T}, ~\beta_{J} = \beta_{C} +\beta_{T},
\end{eqnarray}
with the central-exchange parameters given by
\begin{eqnarray}
\alpha_{C} = 2C_1 + 2 C_2 \rho^{\beta}  +2 C_3\rho^{\gamma}, ~~\beta_{C} = 2C_1^{\prime} + 2 C_2^{\prime} \rho^{\beta}  +2 C_3^{\prime}\rho^{\gamma}.
\end{eqnarray}
As can be seen from Equations~(\ref{Eq:CPot}) and~(\ref{Eq:SOPot}), the coupling constants $b_{\rm IS}$ and $b_{\rm IV}$ appear in both the central potential $U_t$ and the SO potential $\bm{W}_t$. In contrast, the tensor force contributes only to $\bm{W}_t$ and does not affect $U_t$.

\subsection{Charge and Weak Form Factors}

The normalized nuclear form factors, namely, the charge form factor $F_C(q)$ and the weak charge form factor $F_W(q)$, are calculated by folding the point-nucleon form factors $F_t(q)$ and SO current form factors $F_t^{ls}(q)$ ($t=n,p$) with the corresponding intrinsic nucleon electromagnetic $G_{E/M,t}$ and weak $G_{E/M,t}^{(W)}$ form factors~\cite{Reinhard:2021utv}. The point-nucleon and SO current form factors are defined from the local density $\rho_t$ and spin-current density $\bm{J}_t$ as
\begin{eqnarray}
F_t(q) &=& \int d^3\bm{r}\, e^{\mathrm{i}\bm{q}\cdot\bm{r}}\, \rho_t(\bm{r}), \\
F_t^{ls}(q) &=& \int d^3\bm{r}\, e^{\mathrm{i}\bm{q}\cdot\bm{r}}\, \bm{\nabla}\!\cdot\!\bm{J}_t(\bm{r}).
\end{eqnarray}
The charge and weak-charge form factors are then given by
\begin{eqnarray}
F_{C}(q) &=&\frac{1}{Z} \sum_{t=p, n}\left[G_{E, t}(q) F_{t}(q)+G_{M, t}(q) F_{t}^{(l s)}(q)\right],\\
F_{W}(q) &=&\sum_{t=p, n}\frac{\left[G_{E, t}^{(W)}(q) F_{t}(q)+G_{M, t}^{(W)}(q) F_{t}^{(l s)}(q)\right]}{Z Q_{p}^{(W)}+N Q_{n}^{(W)}},
%ZZ: We have highlighted the revised imaginary unit for consistency.
\end{eqnarray}
where $N(Z)$ is the neutron (proton) number, and  $Q_p^{(W)} = 0.0713$ and $Q_n^{(W)}=-0.9888$ are proton and neutron weak charges, respectively.
The electromagnetic form factors $G_{E/M,t}$ are constructed from the isospin-coupled Sachs form factors with the inclusion of the relativistic Darwin correction. Following Ref.~\cite{Reinhard:2021utv}, the weak intrinsic nucleon form factors $G_{E/M,t}^{(W)}$ are obtained from the standard neutral-current decomposition and are expressed in terms of the proton, neutron, and strange electromagnetic form factors (see, e.g., Ref.~\cite{Horowitz:2012we} for details):
\begin{eqnarray}
G_{E/M,n}^{(W)} &=& Q_n^{(W)}G_{E/M,p}+Q_p^{(W)}G_{E/M,n}+Q_n^{(W)}G_{E/M,s}, \\
G_{E/M,p}^{(W)} &=& Q_p^{(W)}G_{E/M,p}+Q_n^{(W)}G_{E/M,n}+Q_n^{(W)}G_{E/M,s},
\end{eqnarray}
where the strange-quark electromagnetic form factors $G_{E/M,s}$ are taken as
\begin{eqnarray}
G_{E,s}(q) &=& \rho_s \frac{\hbar^{2} q^{2}/\left(4 c^{2} m_{N}^{2}\right)}{1+4.97\,\hbar^{2} q^{2}/\left(4 c^{2} m_{N}^{2}\right)},
\end{eqnarray}
\begin{eqnarray}
G_{M,s}(q) &=& \frac{\kappa_s}{\left(1+\dfrac{q^2}{12\kappa_s}r_{M,s}^2\right)^2}.
\end{eqnarray}
Here, $\rho_s=-0.24$~\cite{HAPPEX:2006oqy,Liu:2007yi}, $\kappa_s=-0.017$~\cite{Alexandrou:2019olr}, and $r_{M,s}^2=-0.015~\mathrm{fm}^2$~\cite{Alexandrou:2019olr} are adopted for the strange electric coupling, strange magnetic moment, and strange magnetic radius, respectively. The center-of-mass correction to $F_C$ and $F_W$ is approximately taken into account by renormalizing the nucleon mass $m_N$ to $(1-1/A)m_N$ in the HFB calculation (see, e.g., Ref.~\cite{Bender:2003jk} for details).
%%%%%%%%%%%%%%%%%%%%%%%%%%%%%%%%%%%%%%%%%%
\section{Results}

\subsection{Single-Particle Energy Spectrum of $\rm ^{48}Ca$ and $\rm ^{208}Pb$}

As shown in Ref.~\cite{Yue:2024srj}, {compared with the conventional-IVSO reference eS53, the eS250 functional with $(b_{\rm IS},b_{\rm IV})\approx(160,250)$ MeV$\cdot$fm$^5$ provides a representative strong-IVSO case: it selectively shifts the calculated $\Delta F_{\rm CW}$ of $^{48}$Ca toward the CREX value while leaving the corresponding $^{208}$Pb prediction essentially unchanged and still consistent with PREX-II within the reported $1\sigma$ uncertainty.} For this EDF, the description of ground-state nuclear properties, such as binding energies, charge radii, and spin–orbit splittings of single-particle levels, as well as selected excited-state observables including the giant monopole resonance in $^{208}$Pb and the electric dipole polarizability $\alpha_{\rm D}$, remains in good agreement with experimental data~\cite{Yue:2024srj}. In addition, the pure neutron matter (PNM) equation of state (EOS) predicted by this EDF is consistent with microscopic chiral effective field theory (ChEFT)~\cite{Yue:2024srj}.

\begin{figure}[H]
\centering
\begin{minipage}[b]{1.0\linewidth}
\centering
\includegraphics[width=\linewidth]{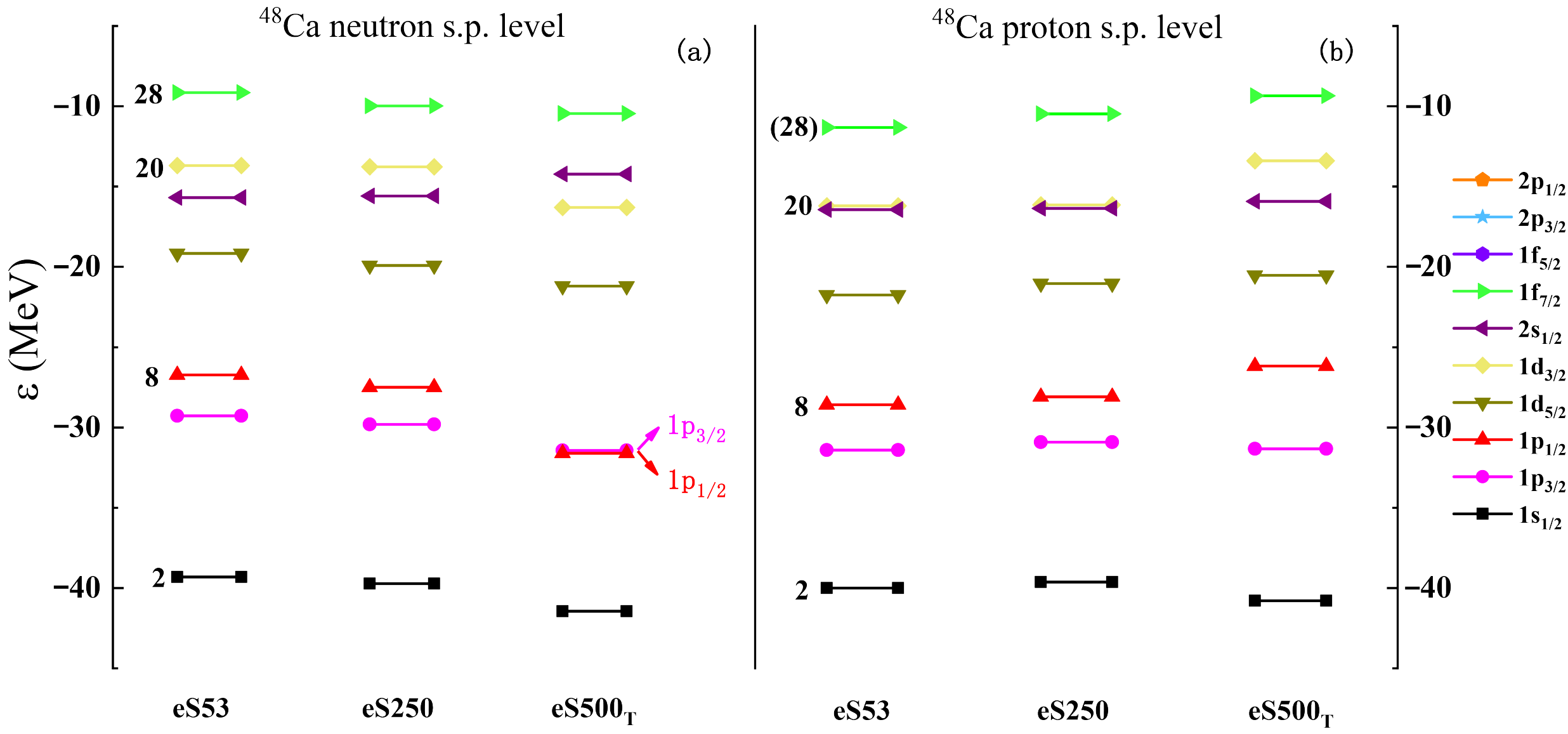}
\end{minipage}
\par\vspace{0.3cm}
\begin{minipage}[b]{1.0\linewidth}
\centering
\includegraphics[width=\linewidth]{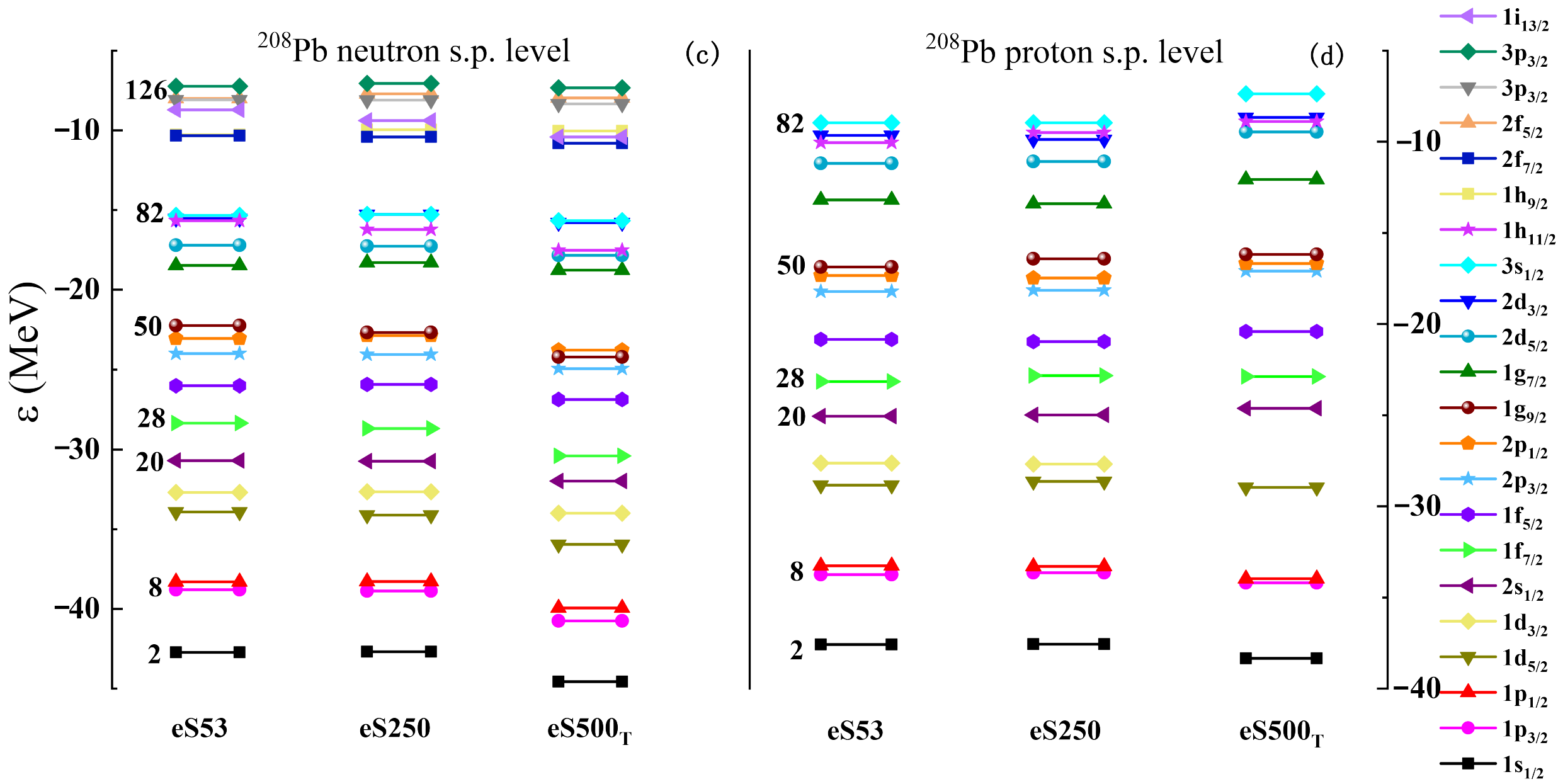}
\end{minipage}
\caption{\textls[-10]{Single-particle energy levels for $^{48}$Ca and $^{208}$Pb. Panel (\textbf{a}) shows neutron levels in $^{48}$Ca; panels (\textbf{b}--\textbf{d}) display proton levels in $^{48}$Ca, neutron levels in $^{208}$Pb, and proton levels in $^{208}$Pb,~respectively.}}
\label{fig:single_particle_levels}
\end{figure}

Moreover, owing to the enhanced IVSO interaction, which further promotes spin–orbit splitting, the single-particle levels and shell structure of many nuclei are affected, with potential implications for magic numbers. A further examination of the single-particle spectra of $^{208}$Pb and $^{48}$Ca, as shown in Figure~\ref{fig:single_particle_levels}, reveals that eS250 reproduces the magic shell closures as well as the conventional IVSO-strength EDF eS53~\cite{Yue:2024srj}. However, when the IVSO interaction is further enhanced using the EDF eS500$_{\rm T}$ with extreme strength from Ref.~\cite{Yue:2024srj}, the single-particle energies in $^{48}$Ca and $^{208}$Pb exhibit certain shifts compared to those obtained with eS53 and eS250. Nevertheless, the magic number structure remains well reproduced, although it slightly violates the usual energy level ordering of the neutron $\rm 1p_{3/2}$ and $\rm 1p_{1/2}$ states in $^{48}$Ca. It is worth emphasizing that when $b_{\rm IV}$ is set to as high as $500$ MeV$\cdot$fm$^5$ as in the eS500$_{\rm T}$, such an extremely large IVSO strength significantly alters the spin–orbit splittings. To partially compensate for these effects, the tensor force is included in the eS500$_{\rm T}$ EDF (see Supplemental Material (SM) in Ref.~\cite{Yue:2024srj}). Moreover, the eS500$_{\rm T}$ fails to reproduce the $\alpha_{\rm D}$ in $^{208}$Pb and violates the PNM EOS from microscopic ChEFT calculations. This indicates that a $b_{\rm IV}$ value of $500$ MeV$\cdot$fm$^5$ is excessively large, whereas $b_{\rm IV}\approx 250$ MeV$\cdot$fm$^5$ represents a more suitable strong-IVSO choice within the adopted EDF family.

\subsection{Sensitivity of $\Delta F_{\rm CW}$ to the IVSO Interaction for $^{48}$Ca, $^{208}$Pb, $^{90}$Zr, and $^{62}$Ni}

\subsubsection{$^{48}$Ca and $^{208}$Pb}

As elucidated in Ref.~\cite{Yue:2024srj}, the sensitivity of $^{48}$Ca to the IVSO interaction originates from the eight neutrons occupying the $\rm 1f_{7/2}$ orbital. This distinction between $^{48}$Ca and $^{208}$Pb can be traced to the IVSO energy term $\frac{b_{\mathrm{IV}}}{2}\tilde{\bm{J}}\cdot\bm{\nabla}\tilde{\rho}$ in Equation (\ref{Eq:Sky-HF}). As shown in Ref.~\cite{Yue:2024srj}, in $^{48}\mathrm{Ca}$, all proton spin–orbit partners are filled, while the neutron $1\mathrm{f}_{7/2}$ orbital is fully occupied and its $1\mathrm{f}_{5/2}$ partner is empty. This configuration produces a large $\tilde{J}$, enabling the IVSO interaction to modify the central mean field and induce global density rearrangements that alter the contribution of every orbital to $\Delta F_{\mathrm{CW}}$. In $^{208}\mathrm{Pb}$, by contrast, the spin–orbit unsaturated orbitals ($1\mathrm{h}_{11/2}$ for protons and $1\mathrm{i}_{13/2}$ for neutrons) yield comparable positive contributions to $J_{\mathrm{p}}$ and $J_{\mathrm{n}}$, resulting in a small $\tilde{J}$ and hence a weak IVSO effect. A detailed orbital-by-orbital analysis of $F_{\rm C}$ and $F_{\rm W}$ for $^{48}$Ca and $^{208}$Pb, comparing the eS53 and eS250 EDFs, can be found in the lower panel of Figure~S5 in the SM of Ref.~\cite{Yue:2024srj}. As shown there, the contributions in $^{48}$Ca exhibit clear differences across all orbitals when varying the IVSO interaction strength, with the most pronounced variation appearing in the $\rm 1d_{3/2}$ rather than $\rm 1f_{7/2}$ orbital---a consequence of the mean-field rearrangement induced by the IVSO interaction. In contrast, the orbital contributions in $^{208}$Pb remain almost unchanged, confirming its insensitivity to the IVSO interaction.
%ZZ: We have highlighted the revised IVSO energy term to match the notation in Equation~(\ref{Eq:Sky-HF}); the unbolded $J$ symbols in the following text denote the corresponding magnitudes.

From this perspective, one can understand why an effective IVSO interaction can change the $\Delta F_{\rm CW}$ predictions for $^{48}$Ca and $^{208}$Pb in different ways within the extended Skyrme EDF framework. If one instead maintains the IVSO interaction at its conventional strength (i.e., $b_{\rm IV} = b_{\rm IS}/3$) and attempts to adjust the tensor force strengths $\alpha_T$ and $\beta_T$, one finds that {the CREX and PREX-II $\Delta F_{\rm CW}$ values cannot be simultaneously described by such tensor-force adjustments}, despite the fact that the tensor force and the IVSO interaction often appear similar in certain respects, such as their ability to modify the ordering of single-particle levels as noted in Ref.~\cite{Yue:2024srj}. The key distinction between the two can be seen from Equations~(\ref{Eq:CPot}) and~(\ref{Eq:SOPot}). The IVSO interaction contributes to both the central potential $U$ and the spin–orbit potential $W$, whereas the tensor force contributes only to the spin–orbit potential $W$. It is precisely this modification of the central potential $U$ by the IVSO interaction that significantly alters the contributions of all orbitals to $F_{\mathrm{C}}$ and $F_{\mathrm{W}}$~\cite{Yue:2024srj}.

\subsubsection{$^{90}$Zr and $^{62}$Ni}

Following the same line of reasoning used in Ref.~\cite{Yue:2024srj} to analyze the contribution of single-particle orbitals to the IVSO term in the energy density (Equation (\ref{Eq:Sky-HF})), one can infer that nuclei with a spin–orbit pattern analogous to $^{48}$Ca, such as $^{90}$Zr, should also exhibit sensitivity to the IVSO interaction in their $\Delta F_{\rm CW}$ values. Conversely, nuclei similar to $^{208}$Pb, such as $^{62}$Ni, are expected to be insensitive to the IVSO interaction. Applying the analysis framework of Ref.~\cite{Yue:2024srj}, one finds that in $^{90}$Zr, the proton orbitals are spin–orbit saturated, whereas the neutron orbitals are spin–orbit unsaturated, with ten neutrons occupying the $\rm 1g_{9/2}$ orbital playing a role analogous to that of the eight neutrons in the $\rm 1f_{7/2}$ orbital of $^{48}$Ca. In contrast, $^{62}$Ni exhibits spin–orbit unsaturation in both proton and neutron channels: eight protons occupy the $\rm 1f_{7/2}$ orbital, and eight neutrons occupy the same orbital. Consequently, the IVSO contribution to the energy density, which is proportional to $\tilde{J}$, becomes negligible.

\begin{figure}[H]
\includegraphics[width=\linewidth]{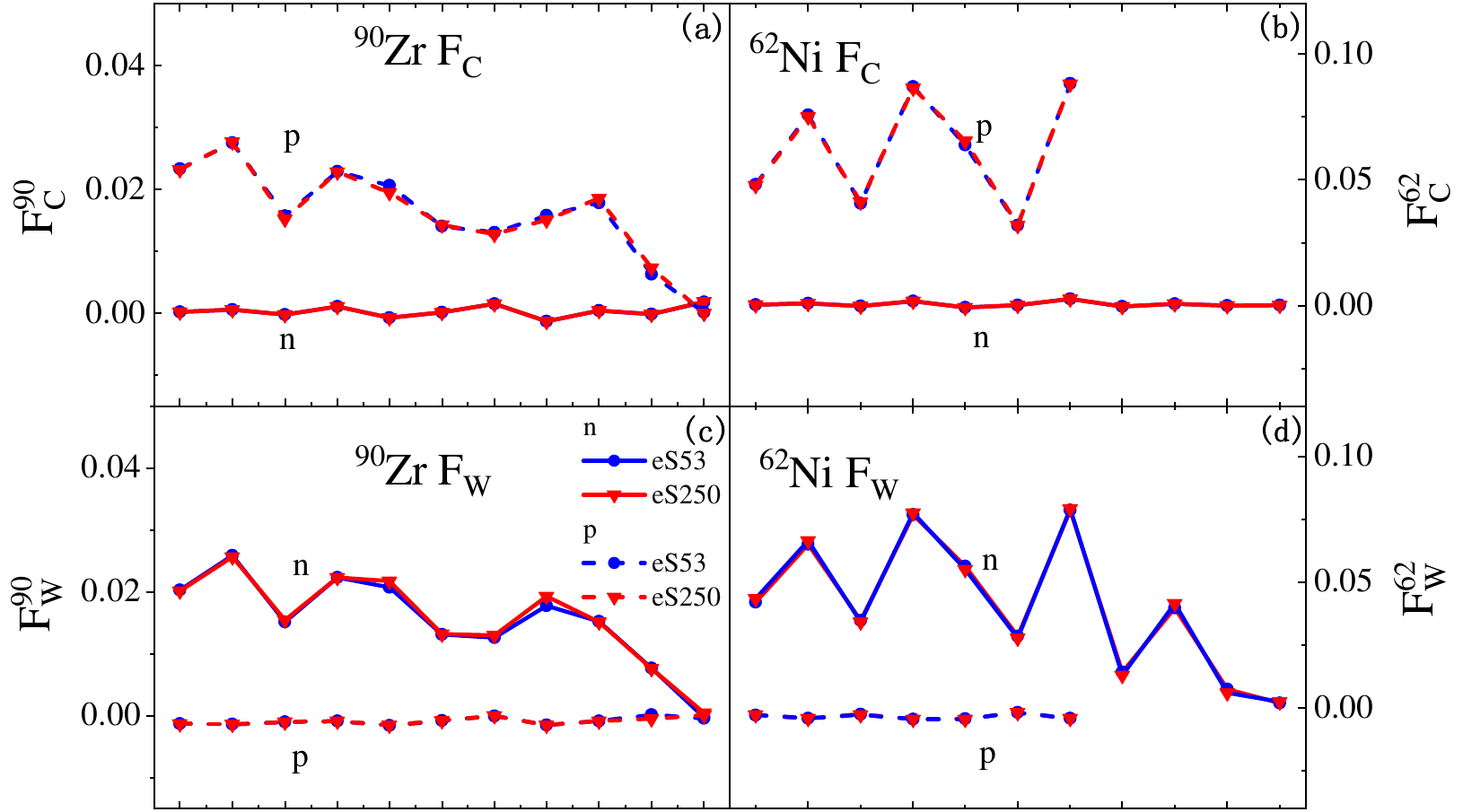}
\caption{Orbital contributions to $F_{\rm C}$ and $F_{\rm W}$ obtained with the eS53 and eS250 EDFs. Panels (\textbf{a},\textbf{c}) show the contributions to $F_{\rm C}$ and $F_{\rm W}$ for $^{90}$Zr, respectively. Panels (\textbf{b},\textbf{d}) show the corresponding results for $^{62}$Ni. The analyzed orbitals (from left to right) include $1{\rm s}_{1/2}$, $1{\rm p}_{3/2}$, $1{\rm p}_{1/2}$, $1{\rm d}_{5/2}$, $1{\rm d}_{3/2}$, $2{\rm s}_{1/2}$, $1{\rm f}_{7/2}$, $1{\rm f}_{5/2}$, $2{\rm p}_{3/2}$, $2{\rm p}_{1/2}$, and $1{\rm g}_{9/2}$, both for $^{90}$Zr and $^{62}$Ni.}
\label{fig:SPFcFw1}
\end{figure}

To illustrate this point, we perform an analysis for $^{90}$Zr and $^{62}$Ni analogous to that presented in the lower panel of Figure~S5 in the SM of Ref.~\cite{Yue:2024srj} for $^{48}$Ca and $^{208}$Pb. As these nuclei are open-shelled, pairing effects must be taken into account. We again compare results obtained with the eS53 and eS250 EDFs, with the proton and neutron pairing strengths determined by fitting the pairing gaps of $^{144}$Sm and $^{120}$Sn~\cite{Bender:2000xk}, respectively, using eS250. The resulting orbital-by-orbital contributions to $F_{\rm C}$ and $F_{\rm W}$ are presented in Figure~\ref{fig:SPFcFw1}a--d. For $^{90}$Zr, as shown in Figure~\ref{fig:SPFcFw1}a,c, each orbital exhibits noticeable differences in its contribution when comparing eS53 and eS250. The most pronounced difference in the contribution to $F_{\rm C}$ appears in the $\rm 1d_{3/2}$ orbital, while for $F_{\rm W}$, the largest difference is observed in the $\rm 1f_{5/2}$ orbital. For $^{62}$Ni, as shown in Figure~\ref{fig:SPFcFw1}b,d, the orbital contributions to both $F_{\rm C}$ and $F_{\rm W}$ are almost identical between eS53 and eS250, confirming our expectation. Meanwhile, we note that the differences in the orbital contributions to $F_{\rm C}$ and $F_{\rm W}$ between eS53 and eS250 are not as pronounced for $^{90}$Zr as they are for $^{48}$Ca, nor are they as negligible for $^{62}$Ni as they are for $^{208}$Pb. This behavior can be attributed, on the one hand, to the diminished influence of the IVSO interaction in heavier nuclei relative to lighter ones and, on the other hand, to the increased number of orbitals contributing in heavier nuclear~systems.

\begin{figure}[H]
\centering
\begin{minipage}[b]{0.9\linewidth}
\centering
\includegraphics[width=\linewidth]{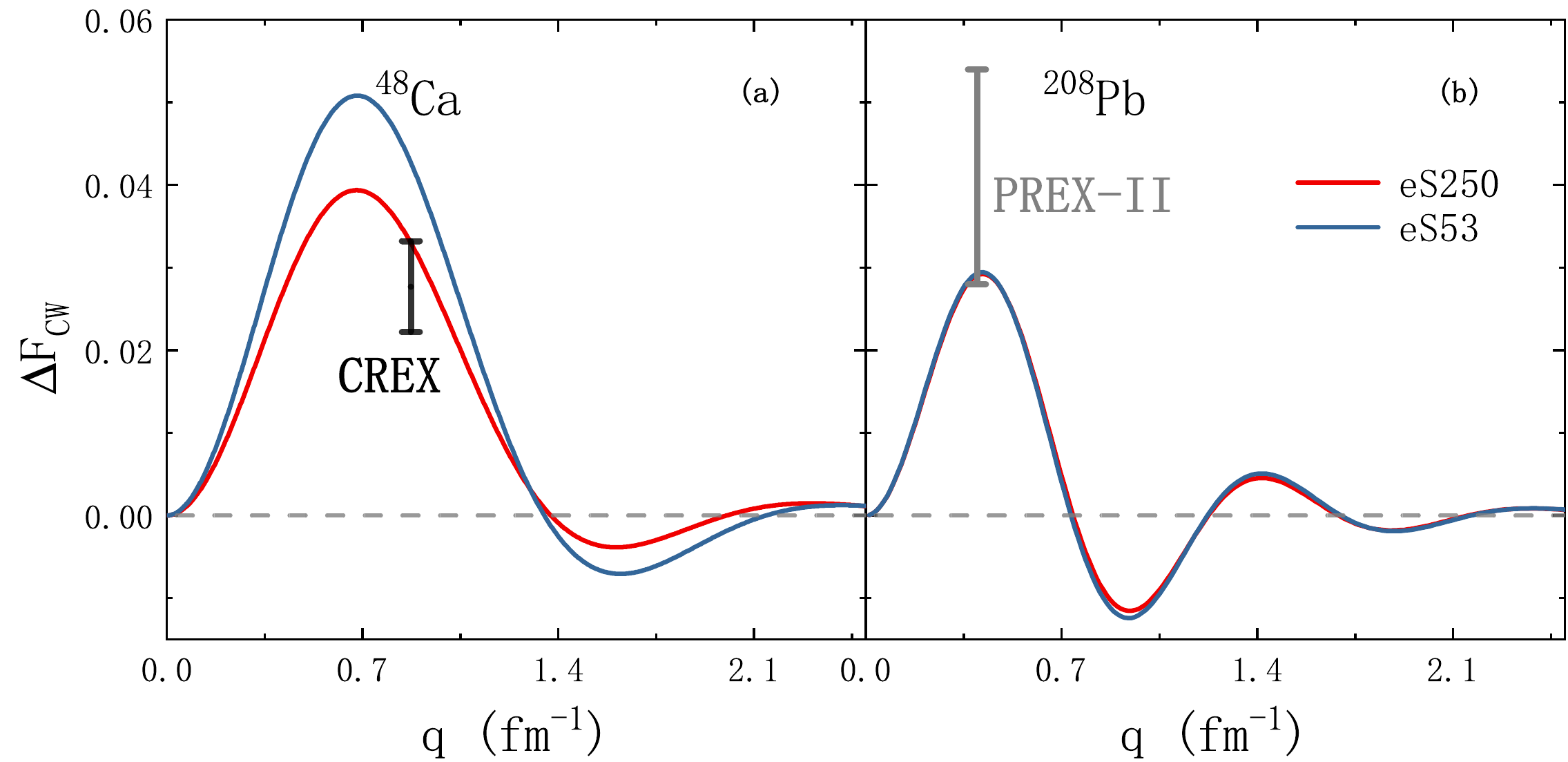}
\end{minipage}
\par\vspace{0.3cm}
\begin{minipage}[b]{0.9\linewidth}
\centering
\includegraphics[width=\linewidth]{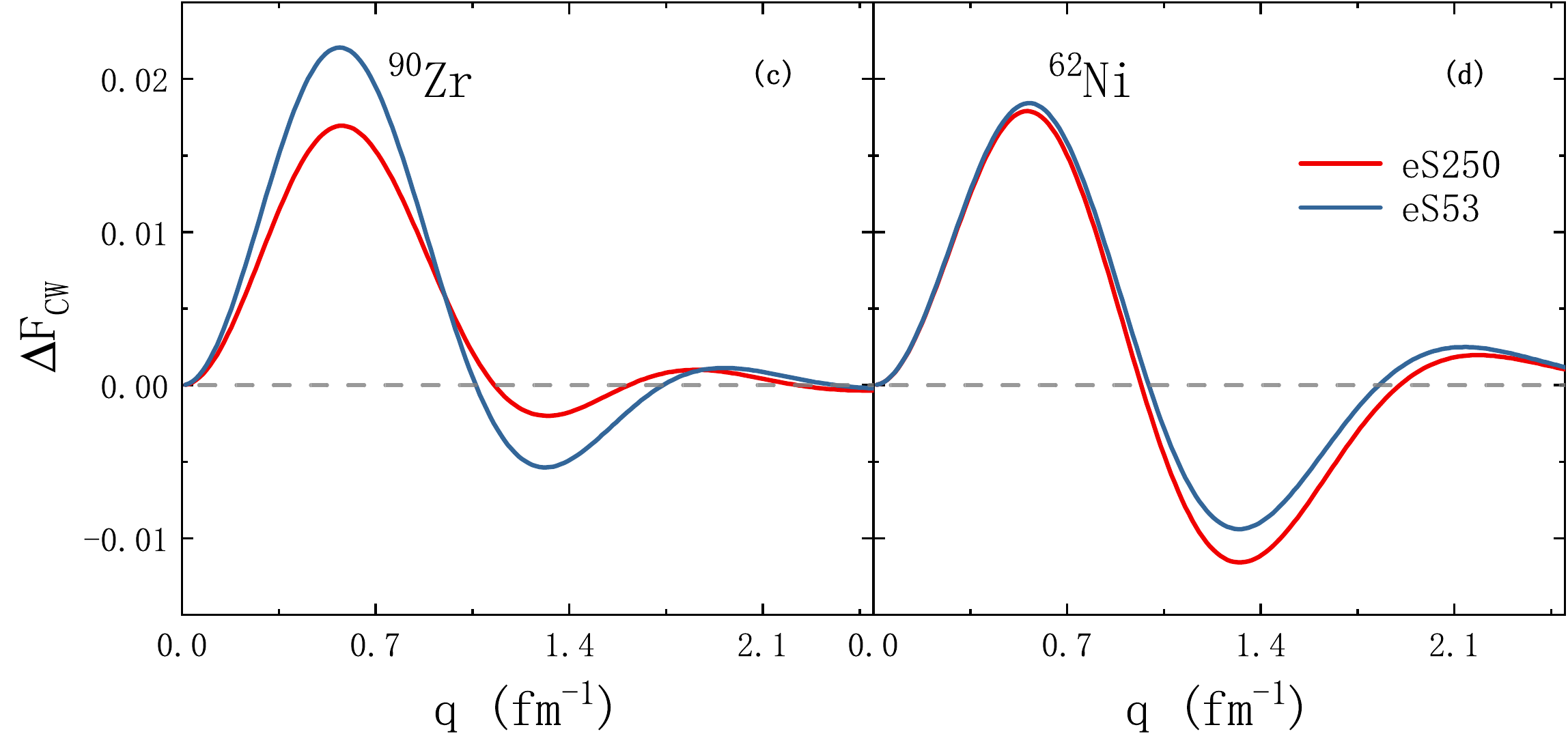}
\end{minipage}
\caption{Momentum-transfer dependence of $\Delta F_{\rm CW}$ for various nuclei. Panels (\textbf{a}--\textbf{d}) show the results for $^{48}$Ca, $^{208}$Pb, $^{90}$Zr, and $^{62}$Ni, respectively, as functions of the momentum transfer $q$ (in fm$^{-1}$).}
\label{fig:DFCWq}
\end{figure}

\textls[-15]{A complementary perspective on this sensitivity is provided by the momentum-transfer dependence of $\Delta F_{\rm CW}$. Figure~\ref{fig:DFCWq} displays this quantity for $^{90}$Zr and $^{62}$Ni in panels (c) and (d)}, respectively, with panels (a) and (b) showing the corresponding results for $^{48}$Ca and $^{208}$Pb for comparison. The comparison reveals that both $^{48}$Ca and $^{90}$Zr exhibit sensitivity to the IVSO interaction. For $^{90}$Zr, this sensitivity is most pronounced at \mbox{$q \approx 0.68$ fm$^{-1}$}. In contrast, $^{62}$Ni and $^{208}$Pb are both insensitive to the IVSO interaction.

The contrasting behavior between the pairs ($^{48}$Ca, $^{90}$Zr) and ($^{208}$Pb, $^{62}$Ni) makes comprehensive experimental measurements on these nuclei particularly meaningful and important. $^{90}$Zr is a stable natural isotope of zirconium with an abundance of 51.45\% in nature~\cite{nndc_nudat}, while $^{62}$Ni is likewise a stable natural isotope of nickel with an abundance of 3.63\%~\cite{nndc_nudat}. Both are suitable for experimental measurements. Owing to the insensitivity of $^{208}$Pb and $^{62}$Ni to the IVSO interaction, parity-violating electron scattering (PVES) experiments measuring $\Delta F_{\rm CW}$ (or the neutron skin thickness) on these two nuclei would provide a combined constraint on the symmetry energy that is more precise than that obtained from PREX-II alone. Meanwhile, PVES measurements on $^{90}$Zr, when combined with the CREX results, would enable a more accurate determination of the IVSO interaction strength $b_{\rm IV}$. Such measurements can be realistically performed at existing and upcoming facilities. The MESA accelerator in Mainz, with its planned PVES program MREX~\cite{Schlimme:2024eky}, is well suited for precision measurements on nuclei including $^{90}$Zr and $^{62}$Ni as well as $^{48}$Ca and $^{208}$Pb. Complementary measurements at JLab, building on the successful PREX-II and CREX experiments, are also feasible, as demonstrated by recent proposals and feasibility studies~\cite{Esser:2025deh,Gal:2024rcg}. This would have a positive and significant impact on deepening our understanding of both the symmetry energy and the IVSO interaction.

%\vspace{-6pt}

\subsection{Neutron Skin $R_{\rm np}$ of $^{48}$Ca, $^{208}$Pb, $^{90}$Zr, $^{62}$Ni, $^{22}$O and $^{24}$O}

Compared to $\Delta F_{\rm CW}$, the neutron skin thickness $R_{\rm np}$ is a more intuitive and frequently discussed quantity, which, like $\Delta F_{\rm CW}$, characterizes the overall difference between neutron and proton distributions. Beyond $^{48}$Ca and $^{90}$Zr, following the analysis in Ref.~\cite{Yue:2024srj}, one can readily identify $^{22}$O and $^{24}$O as nuclei that are also sensitive to the IVSO interaction, with evidence supporting their double magic nature~\cite{Ozawa:2000gj,Becheva:2006zz}. Table~\ref{tab:Nskin} presents the calculated $R_{\rm np}$ values for $^{48}$Ca, $^{208}$Pb, $^{90}$Zr, $^{62}$Ni, $^{22}$O, and $^{24}$O obtained with the eS53 and eS250 EDFs. For $^{208}$Pb and $^{62}$Ni, the results from the two EDFs, are nearly identical. In contrast, for $^{48}$Ca, $^{90}$Zr, $^{22}$O, and $^{24}$O, the eS250 predictions are smaller than those of eS53, reflecting a behavior similar to that observed for $\Delta F_{\rm CW}$. It is worth noting that the charge-exchange reaction measurement by Ref.~\cite{Yako:2006gz} reported $\Delta R_{\rm np} = 0.07 \pm 0.04$ fm for $^{90}$Zr, with the central value of 0.07 fm lying between the eS250 and eS53 predictions, indicating the need for more precise experimental determinations. Furthermore, future measurements of $R_{\rm np}$ for $^{22}$O and $^{24}$O, e.g., by charge-changing cross sections~\cite{Super-FRS:2024eei,Hasan:2024lvp}, would be highly desirable to validate the predictions of the present work.

\begin{table}[H]
\caption{Neutron skin $R_{\rm np}$ (in fm) of $^{48}$Ca, $^{208}$Pb, $^{90}$Zr, $^{62}$Ni, $^{22}$O and $^{24}$O with eS53 and eS250. \label{tab:Nskin}}
\begin{tabularx}{\textwidth}{LCC}
\toprule
\textbf{Nucleus} & \textbf{eS53} & \textbf{eS250} \\
\midrule
$^{48}$Ca & 0.179 & 0.129 \\
$^{208}$Pb & 0.199 & 0.195 \\
$^{90}$Zr & 0.083 & 0.053 \\
$^{62}$Ni & 0.096 & 0.093 \\
$^{22}$O  & 0.367 & 0.290 \\
$^{24}$O  & 0.510 & 0.430 \\
\bottomrule
\end{tabularx}
\end{table}
\section{Conclusions}

In this work, we have systematically investigated the sensitivity of the charge-weak form factor difference $\Delta F_{\rm CW}$ to the IVSO interaction in $^{48}$Ca, $^{208}$Pb, $^{90}$Zr, and $^{62}$Ni using the extended Skyrme EDFs eS53 and eS250. {Relative to eS53, the strong-IVSO functional eS250 with $(b_{\rm IS},b_{\rm IV})\approx(160,250)$ MeV$\cdot$fm$^5$ selectively shifts the calculated $\Delta F_{\rm CW}$ of $^{48}$Ca toward the CREX value while leaving the corresponding prediction for $^{208}$Pb essentially unchanged~\cite{Yue:2024srj}. At the same time, eS250 preserves the established magic shell closures in $^{48}$Ca and $^{208}$Pb and maintains a good description of global nuclear properties.}

By examining the orbital-by-orbital contributions to $F_{\rm C}$ and $F_{\rm W}$, we demonstrate that the sensitivity of $^{48}$Ca originates from the eight $\rm 1f_{7/2}$ neutrons, which modify the central mean-field potential and thereby influence all orbitals. A similar mechanism is identified in $^{90}$Zr, where ten $\rm 1g_{9/2}$ neutrons play an analogous role, leading to a pronounced IVSO sensitivity in its $\Delta F_{\rm CW}$ and neutron skin thickness $R_{\rm np}$. In contrast, $^{208}$Pb and $^{62}$Ni, characterized by near-cancellation of the isovector spin–orbit density $J_{\rm n}-J_{\rm p}$, exhibit little sensitivity to the IVSO interaction.

This structural distinction between the pairs ($^{48}$Ca, $^{90}$Zr) and ($^{208}$Pb, $^{62}$Ni) offers a controlled experimental strategy for disentangling the IVSO interaction from the density dependence of the symmetry energy. The natural abundances and experimental accessibility of $^{90}$Zr and $^{62}$Ni make them ideal candidates for future PVES measurements. Such measurements could be performed at MREX and JLab. Once available, they can be combined with existing PREX-II and CREX observables to help separate EDF-level IVSO effects from symmetry-energy effects and thereby constrain both the density dependence of the symmetry energy and the IVSO strength $b_{\rm IV}$.

Furthermore, our predictions for the neutron skin thickness in $^{22}$O and $^{24}$O, which could be measured by charge-changing cross sections, suggest that these neutron-rich oxygen isotopes are also sensitive to the IVSO interaction, offering additional testing grounds for the proposed strong IVSO scenario.
Overall, our results show that the sensitivity of $\Delta F_{\rm CW}$ to the IVSO interaction is strongly nucleus-dependent and is closely tied to the underlying spin–orbit saturation pattern. In this respect, PVES measurements across different nuclei offer a valuable way to constrain the isovector spin–orbit channel in nuclear EDFs.
\vspace{6pt}
\authorcontributions{T.-G.Y. performed the theoretical calculations.
Z.Z. and L.-W.C. supervised the theoretical
calculations. All authors jointly analyzed the data and
contributed to the interpretation of the results and to the
writing of the manuscript. All
 authors have read and agreed to the published version of the manuscript.}

\funding{This work was supported in part by the National Natural Science Foundation of China under Grant Nos.
  12235010 and 12575137, the National SKA Program of China No.~2020SKA0120300, and the Science and Technology Commission of Shanghai Municipality under Grant No. 23JC1402700.}

\dataavailability{All the data supporting the findings in this work are
available within the manuscript, and any additional data
are available from the corresponding authors upon rea-
sonable request.}

\acknowledgments{The authors would like to thank Wei-Zhou Jiang, Yifei Niu, Yifeng Sun, and Yu-Min Zhao for useful discussions.}

\conflictsofinterest{The authors declare that they have no conflicts of interest.}

\begin{adjustwidth}{-\extralength}{0cm}

\reftitle{References}

\PublishersNote{}
%\isPreprints{}{% This command is only used for ``preprints''.
\end{adjustwidth}
\end{document}